# Origin of anomalous anharmonic lattice dynamics of lead telluride

Takuma Shiga[1], Takuru Murakami[1], Takuma Hori[1],

Olivier Delaire[2], and Junichiro Shiomi[1,3,*],

[1]*Department of Mechanical Engineering, The University of Tokyo, 7-3-1 Hongo,*

*Bunkyo, Tokyo, 113-8656, Japan*

[2]*Neutron Scattering Science Division, Oak Ridge National Laboratory, 1 Bethel Valley*

*Road, Oak Ridge, Tenessee 37831, USA*

[3]*Japan Science and Technology Agency, PRESTO, 4-1-8 Honcho, Kawaguchi, Saitama,*

*332-0012, Japan*

**Abstract**

The origin of the anomalous anharmonic lattice dynamics of lead telluride is investigated using molecular dynamics simulations with interatomic force constants (IFCs) up to quartic terms obtained from first principles. The calculations reproduce the peak asymmetry of the radial distribution functions and the double peaks of transverse optical phonon previously observed with neutron diffraction and scattering experiments. They are identified to be due to the extremely large nearest-neighbor cubic IFCs in the [100] direction. The outstanding strength of the nearest-neighbor cubic IFCs relative to the longer-range ones explains the reason why the distortion in the radial distribution function is local.

Knowledge of the lattice anharmonicity, which is a source of intrinsic lattice thermal resistance, thermal expansion, and temperature dependence of elastic constant,[1,2] has recently become accessible with development in numerical and experimental tools to probe anharmonic lattice dynamics. Lead telluride (PbTe) has been a popular material of choice for studying lattice anharmonicity because it exhibits an intrinsically low lattice thermal conductivity (2.0~2.2 $Wm^{-1}K^{-1}$ at room temperature[3]), which is beneficial for thermoelectric applications.[4-8] Calculations of mode-dependent phonon scattering rates have revealed that heat conduction by the longitudinal acoustic (LA) phonons is significantly inhibited by scattering with the strongly anharmonic transverse optical (TO) phonons.[7]

Further anomalous anharmonic characteristics of PbTe lattice dynamics have been experimentally observed in the radial distribution function from neutron diffraction[9] and the inelastic neutron scattering (INS) spectra.[10] The nearest-neighbor peak in the pair distribution function (PDF) obtained from the neutron diffraction experiments was found to deviate from a Gaussian distribution function with the deviation magnitude increasing with temperature, while globally maintaining the rock-salt structure.[9] In addition, it was emphasized that the PDF for the nearest neighboring Pb and Te atoms along [100] direction is surprisingly broad,[9] although it has long been known that PbTe exhibits large atomic displacement parameters (ADPs). More recent synchrotron x-ray powder diffraction experiment confirmed these large ADPs, and interpreted the synchrotron measurements in terms of Pb atoms that are off-centered from their equilibrium positions.[11] Interestingly, the magnitude of the deviation was found to increase with temperature unlike the ferroelectric materials, such as perovskites, where the transition to paraelectric (displacive phase transition) occurs with increasing temperature.[1] However, models presented in Refs. 9 and 11 based on interpretations of the PDF and ADP from diffraction experiments reach a different conclusion than studies of bond lengths based on measurements of the extended x-ray absorption fine structure (EXAFS), which concluded that Pb atoms are not off-centered, but undergo vibrations of large amplitude.[12, 13] In addition, the asymmetry in PDF peaks has been reproduced by *ab initio* molecular dynamics simulations[13, 14], but these simulations also showed that there are no appreciable off-centerings nor local or global spontaneously broken symmetries at finite temperatures.[13, 15]

The strong anharmonicity of PbTe also manifests in the phonon dispersion relations. While the LA and TO phonon branches calculated using the harmonic theory[4, 6, 7] cross each other along the [100] symmetry line, INS experiments have identified that the LA-TO crossing is avoided (avoided crossing)[10] at finite temperature, as has been seen in perovskites.[10, 16] The experiment also found a peak emerging at the zone center with a frequency that is different from TO and longitudinal optical (LO) phonons, resulting in a *double peak* in the INS line shape. The magnitude of the double peak increases with temperature, suggesting that it originates from the lattice anharmonicity.

The temperature dependence of the INS spectrum has been studied based on a perturbation theory.[9] The self-consistent *ab initio* lattice dynamics (SCAILD)[17] method was used to calculate the temperature dependence of TO-phonon frequency at the zone center by modeling the anharmonicity using the quasi-harmonic theory.[2] The hardening of the TO-phonon frequency with increasing temperature was qualitatively reproduced, although the emergence of the double peak features were not captured because the perturbation theory does not account for changes in the eigenstates.

In contrast, molecular dynamics (MD) method can readily include changes in the eigenstates arising from lattice anharmonicity. In particular, the usability of the classical MD method on studying low Debye-temperature materials such as PbTe has been greatly improved with the recent development of non-empirical force fields consisting of accurate anharmonic interatomic force constants (IFCs) obtained from first principles. The MD method with anharmonic IFCs has been shown to accurately reproduce lattice thermal conductivity of pure and alloyed crystals.[18, 19]

In this work, we have investigated the peak asymmetry in radial distribution function, avoided-crossing, and double peak in PbTe crystal using the classical MD method with first-principles-based anharmonic IFCs. IFCs are defined as the Taylor expansion coefficient of the potential energy ($V$) or interatomic force ($F$) with respect to the atomic displacement ($u$) around the equilibrium coordination:

$$F_i^\alpha = -\frac{\partial V}{\partial u_i^\alpha} = -\sum_{j,\beta} \Phi_{ij}^{\alpha\beta} u_j^\beta - \frac{1}{2!}\sum_{jk,\beta\gamma} \Psi_{ijk}^{\alpha\beta\gamma} u_j^\beta u_k^\gamma - \frac{1}{3!}\sum_{jkl,\beta\gamma\delta} \mathrm{X}_{ijkl}^{\alpha\beta\gamma\delta} u_j^\beta u_k^\gamma u_l^\delta, \quad (1)$$

where $\Phi$, $\Psi$, and $\mathrm{X}$ are harmonic, cubic, and quartic anharmonic IFCs, respectively. The sub- and superscripts $i$, $j$, $k$, and $l$ are atomic indices, and $\alpha$, $\beta$, $\gamma$, and $\delta$ are Cartesian coordinates. The IFCs up to quartic term have been calculated using the *real-space displacement* method[20] with the Hellman-Feynman forces obtained by density-functional-theory calculations. The harmonic, cubic, and quartic IFCs were calculated considering the interaction up to sixth, first, and first neighboring atoms, respectively. More details of the calculation are described in the previous reports.[7, 18] Even with the care to stabilize the MD simulation with quartic terms and random displacements described in Ref. 18, we could not realize stable MD simulations above 300 K, and thus, this letter reports results at temperatures smaller than 300 K.

We first computed the radial distribution function, $g(r)$, from the MD trajectories. MD simulations were typically performed for $4 \times 4 \times 4$ cubic supercell with 512 atoms. In each simulation, the trajectories were recorded for 600 ps with constant energy (i.e. microcanonical ensemble) after equilibration at particular temperature. The time step was set to 2.0 fs. For each temperature, ten MD simulations with different initial conditions were performed and the calculated $g(r)$ was ensemble-averaged.

As shown in Fig. 1(a), the magnitude and width of the peak in $g(r)$ decreases and increases, respectively, with increasing temperature. The first peak of $g(r)$, denoted hereafter as $g^{1stPb-Te}(r)$, corresponding to the distance between the nearest-neighboring Pb and Te atoms, exhibits asymmetric non-Gaussian profile. To clarify the trend, we used the Levenberg-Marquardt algorithm[21] to fit the $g^{1stPb-Te}(r)$ profiles by Gaussian distribution functions, as plotted in Figs. 1(b) and 1(c). It can be seen that $g^{1stPb-Te}(r)$ significantly deviates from the Gaussian distribution at higher temperature (250 K), which agrees well with the *ab initio* MD simulations[13, 14] and the experiment.[9] Here we confirmed that, despite the distortion of $g^{1stPb-Te}(r)$ from the Gaussian distribution, the average-position of Pb atom is not off-centered, which is consistent with the *ab inito* MD simulation[13] and the EXAFS experiment.[12]

To quantify the temperature dependence of the peak asymmetry in $g^{1stPb\text{-}Te}(r)$, we calculated Skewness and Kurtosis through third ($n=3$) and fourth ($n=4$) order momentum of the distribution functions.

$$\text{Skewness} = \frac{m_3}{\sigma^3}, \quad \text{Kurtosis} = \frac{m_4}{\sigma^4} - 3, \quad m_n = \frac{\int_{-\infty}^{\infty} dr (r-\mu)^n g(r)}{\int_{-\infty}^{\infty} dr g(r)}, \tag{2}$$

where $\mu$ and $\sigma$ denote the expectation and variance of $g^{1stPb\text{-}Te}(r)$, respectively. Skewness and Kurtosis respectively quantify the asymmetry and flatness of the profile. Kurtosis is zero if the profile is Gaussian. Figures 1(d) and 1(e) show the temperature dependence of Skewness and Kurtosis of $g^{1stPb\text{-}Te}(r)$. For comparison, the values obtained from the radial distribution function of nearest-neighboring Pb (Te) and Pb (Te) atoms are also plotted. For the entire temperature range, the calculated Kurtosis of $g^{1stPb\text{-}Pb}(r)$ and $g^{1stTe\text{-}Te}(r)$ are nearly zero, indicating that their profiles are Gaussian. In contrast, the deviation of $g^{1stPb\text{-}Te}(r)$ from the Gaussian distribution monotonically increases with temperature. Similarly, the positive value of Skewness of $g^{1stPb\text{-}Te}(r)$ increases with temperature, indicating that the anharmonicity skews the profile towards small distance ($r$). The linear trend of Skewness is consistent with that of the extent of asymmetry measured in the experiments, although the dependence in the experiments became week as temperature exceeded 225 K.[9] In addition, it is interesting to note that Kurtosis of $g^{1stPb\text{-}Te}(r)$ exhibits transition-like drastic increase above 200 K, which may be worth more detailed investigation in the future.

We next calculated the dynamical structure factor,[22, 23] $S(\mathbf{Q},\omega)$ for comparison with the INS experiments.[10, 24]

$$S(\mathbf{Q},\omega) = \frac{1}{2\pi N}\sum_{j,k} b_j b_k \int_{-\infty}^{\infty} dt \langle e^{-i\mathbf{Q}\cdot\mathbf{r}_j(0)} e^{i\mathbf{Q}\cdot\mathbf{r}_k(t)}\rangle, \tag{3}$$

where $\mathbf{Q}$, $\omega$, and $N$ are wavevector, frequency, and total number of atoms in the system. The scattering length of $j$th atom $b_j$ was set to $9.405 \times 10^{-15}$ and $5.800 \times 10^{-15}$ m for Pb and Te atoms, respectively.[23] The wavevector $\mathbf{Q}$ is sum of the Brillouin zone wavevector $\mathbf{Q}_{BZ}$ and the scanning wavevector $\mathbf{q}$ ($\mathbf{Q}=\mathbf{Q}_{BZ}+\mathbf{q}$). We chose $\mathbf{Q}_{BZ}$ of (113) and (002) Brillouin zones (BZs) following the experiments.[10] For the $S(\mathbf{Q},\omega)$ calculation, we performed MD simulations for a $10 \times 10 \times 10$ cubic supercell with 8,000 atoms to ensure the large enough reciprocal lattice points in the Brillouin zone. In addition, ten 2.0 ns-long MD simulations were performed and ensemble-averaged to reduce the noise in $S(\mathbf{Q},\omega)$.

Figure 2 shows $S(\mathbf{Q},\omega)$ calculated along the [001] symmetry line. White solid lines are phonon dispersion relations obtained by the harmonic lattice dynamics calculations. Here it is noted that the LO-TO splitting around the zone center[7,19] is absent because long-range interactions due to the ionic charge were not incorporated in the current MD simulations for simplicity. At low temperature (50 K), as shown Fig. 2(a), peak positions of $S(\mathbf{Q},\omega)$ in (113) BZ falls on the harmonic phonon dispersion relations except for the TO phonon modes around the zone center. Even at higher temperature (250 K), the $S(\mathbf{Q},\omega)$ of longitudinal phonon modes agree with the phonon dispersion relations as shown in Fig. 2(b). On the other hand, transverse acoustic (TA) phonons become softer near the zone boundary, and the transverse optical phonons become harder throughout the BZ. The hardening of TO phonons is consistent with the consequence of the avoided crossing observed in the experiments,[10] however, the extent of hardening here was not enough to avoid the crossing. While peaks in $S(\mathbf{Q},\omega)$ for all the modes become broader as temperature increases due to phonon scattering, the broadening and deformation of the optical phonon modes around the zone center is extraordinary, resulting in the spectrum completely different from the harmonic calculations. The *V shape* in the $S(\mathbf{Q},\omega)$ spectra around the zone center is in excellent agreement with the INS experiment.[10] The branch assignment of theses optical phonon modes can be done by moving the BZ from (113) to (002), where transverse modes in theory should be invisible. The absence of the extraordinarily broadened and deformed peaks in Fig. 2(c) indicates that the peaks are TO phonon modes.

Figure 3 (a) shows more detail pictures of the temperature dependence of the TO phonon modes at the (113) BZ center. The presence of double peak can be clearly identified. The frequency and line width of both peaks strongly depend on temperature, as observed in the INS experiments.[10, 24] The two peaks were fitted by Gaussian distribution functions using the Levenberg-Marquardt algorithm[21] and the obtained temperature dependences of the peak frequencies are plotted in Fig. 3(b). The frequency of the second peak increases with temperature, and the magnitude of change is larger than that of the first peak, which is consistent with the INS experiments.[10, 24] Although there are still moderate discrepancies between the calculation and the experiments such as the sign of the slope of the first-peak frequency peak around 100 K, in overall, the classical MD simulations appear to capture the qualitative features of $S(\mathbf{Q},\omega)$ reasonably well.

We now identify which IFCs are responsible for the peak asymmetry in $g^{\text{1stPb-Te}}(r)$ and the double peak. It is intuitive to investigate the cubic anharmonic IFCs because they are asymmetry with respect to the atomic displacement. Considering the crystal symmetries of PbTe, there are in total six irreducible cubic IFCs, $\Psi_{ijk}^{\alpha\beta\gamma}$, as listed in Tab. 1. The corresponding atomic displacements are shown in Fig. 4(a). The table shows a surprising feature that the cubic anharmonic IFCs along the [100] direction ($\Psi_{\text{PbTeTe}}^{zzz}$ and $\Psi_{\text{PbPbTe}}^{zzz}$) are more than 50 times larger than the others. We have also checked cubic IFCs up to third neighbors and confirmed that they are smaller than the nearest-neighbor ones. To check the impact of the cubic IFCs on the line shape of the zone-center TO phonons, we performed additional simulations at 200 K using $\Psi_{\text{PbTeTe}}^{zzz}$ and $\Psi_{\text{PbPbTe}}^{zzz}$ with reduced magnitudes. The magnitudes were reduced in steps by scaling $\Psi_{\text{PbTeTe}}^{zzz}$ and $\Psi_{\text{PbPbTe}}^{zzz}$ with a factor $\gamma$ less than unity. As shown in Fig. 4(b), the magnitude of the double peak decreases as $\gamma$ decreases, and completely disappears as $\gamma$ approaches zero. Note that, in the case of $\gamma=0$, the peak frequency 5 meV is 25% larger than that of the harmonic calculation due to the anhamonicity of the other IFCs including the quartic ones. Similar parameter study was also carried out for the radial distribution function and the distribution was found to become Gaussian by reducing the value of $\gamma$ to zero. These results prove that it is indeed these nearest-neighbor cubic IFCs in the [100] direction that causes both the peak asymmetry in $g^{\text{1stPb-Te}}(r)$ and the double peak. In addition, the fact that it is the nearest-neighbor IFCs explains why the distortion is limited to the nearest-neighbor peaks in the radial distribution function.

In summary, we have performed classical MD simulations of PbTe crystal using

IFCs up to quartic terms obtained from first principles. The radial distribution functions and dynamical structural factors calculated from the MD simulations qualitatively reproduces anomalous anharmonic-lattice-dynamics characteristics of PbTe such as peak asymmetry in $g^{1stPb\text{-}Te}(r)$ and double peaks observed in neutron detraction and scattering experiments, respectively. The anharmonic behavior was identified to be due to the extremely large nearest neighbor cubic IFCs along the [100] direction. The outstanding strength of the nearest-neighbor cubic IFCs to the longer-range ones also explains the reason why the distortion of the radial distribution function is limited to short range. This letter also demonstrates the usability of MD simulations with anharmonic IFCs to identify the origin of anharmonic lattice dynamics, which may become useful to understand and engineer old and new materials with preferred anharmonic thermodynamic properties.

**Acknowledgements**

The authors are grateful for discussions with Dr. Gang Chen and Dr. Keivan Esfarjani. This work was partially supported by Research Fellowships of the Japan Society for the Promotion of Science for Young Scientists, Japan Science and Technology Agency PRESTO, and KAKENHI 23760178.

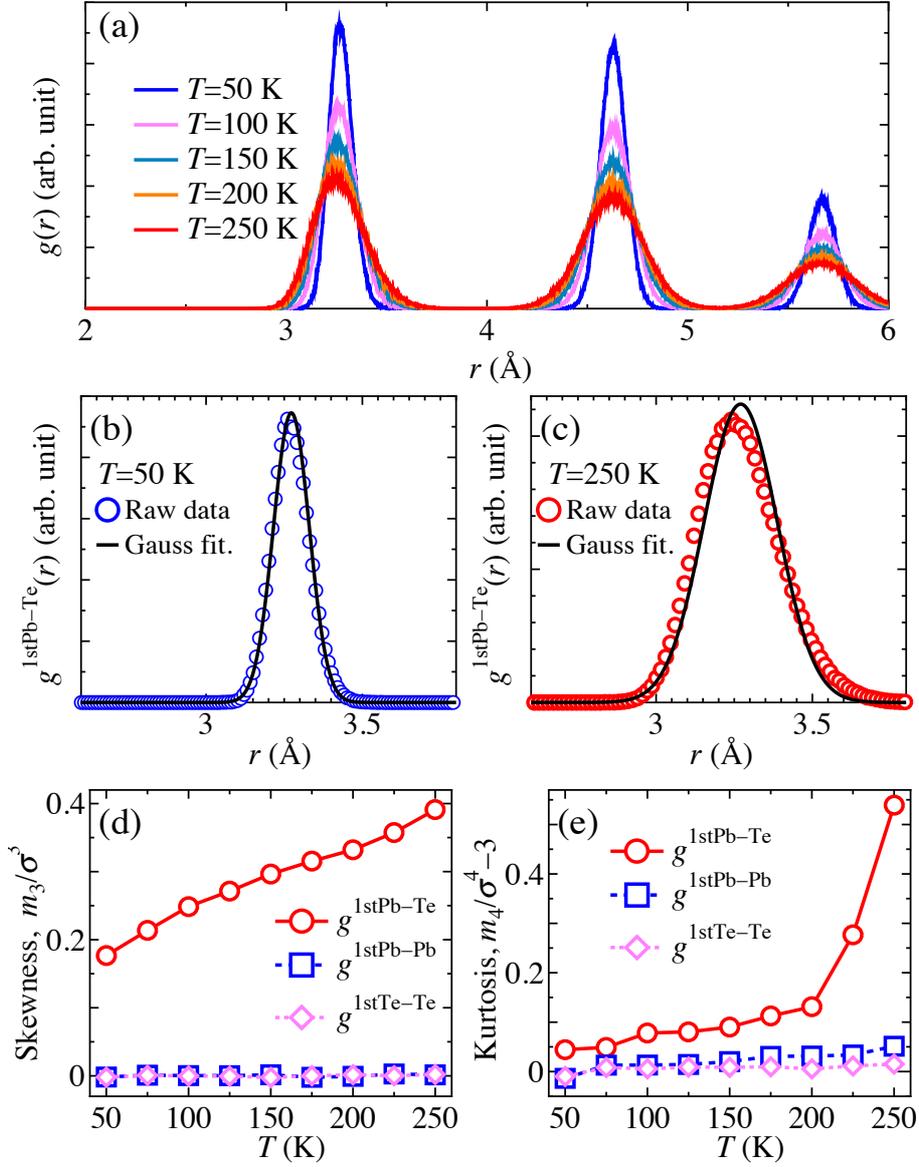

**Fig. 1.** (Color online) (a) Radial distribution function, $g(r)$, of pure PbTe crystal obtained by MD simulations at five different temperatures ($T$=50, 100, 150, 200, and 250 K). (b) and (c) $g(r)$ of the nearest neighbor Pb-Te, $g^{1stPb\text{-}Te}(r)$, at $T$=50 and 250 K, respectively. Open circles are the raw data, and the solid lines denote the Gaussian distribution functions fitted using the Levenberg-Marquardt algorithm.[21] (d) and (c) temperature-dependent Kurtosis and Skewness of $g^{1stPb\text{-}Te}(r)$, $g^{1stPb\text{-}Pb}(r)$, and $g^{1stTe\text{-}Te}(r)$.

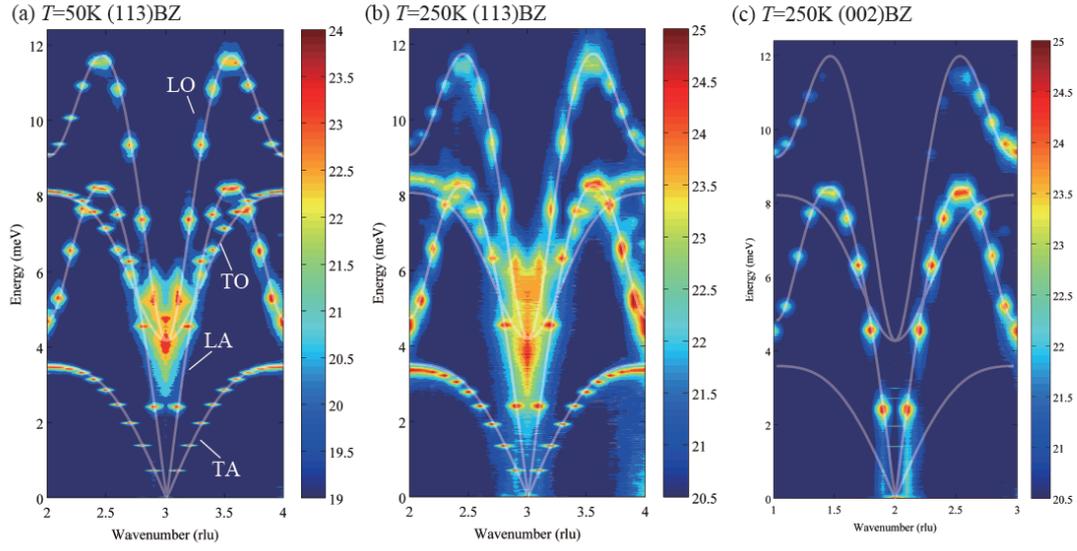

**Fig. 2.** (Color online) Logarithmic contour plots of dynamical structure factor, $S(\mathbf{Q},\omega)$, along [001] symmetry line. (a) and (b) $S(\mathbf{Q},\omega)$ spectra in (113) BZ at $T$=50 and 250 K, respectively. (c) $S(\mathbf{Q},\omega)$ spectra in (002) BZ at $T$=250 K. White solid lines denote the phonon dispersion relations obtained by the harmonic lattice dynamics calculations.

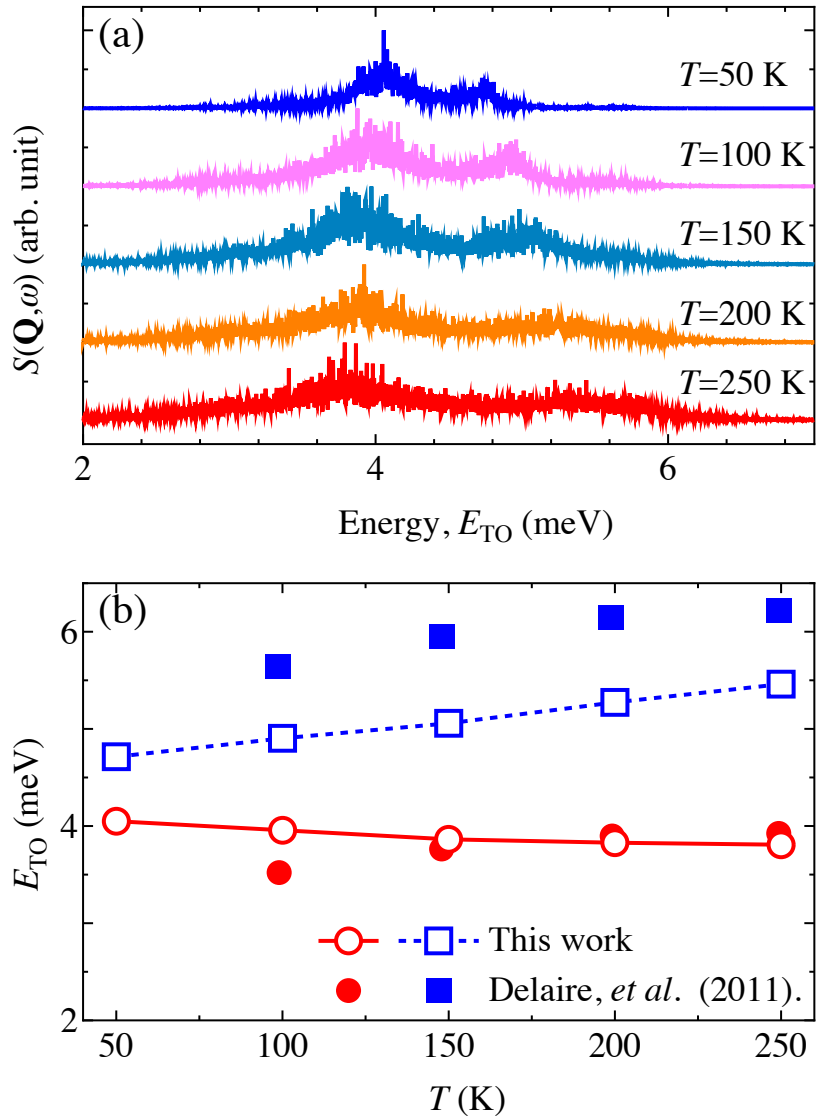

**Fig. 3.** (Color online) (a) Temperature dependence of $S(\mathbf{Q},\omega)$ at (113) BZ calculated by MD simulations. (b) Temperature dependence of frequencies of the first- and second-peaks fitted by Gaussian using the Levenberg-Marquardt algorithm.[21] Filled circles and squares are the values from the INS experiment.[10]

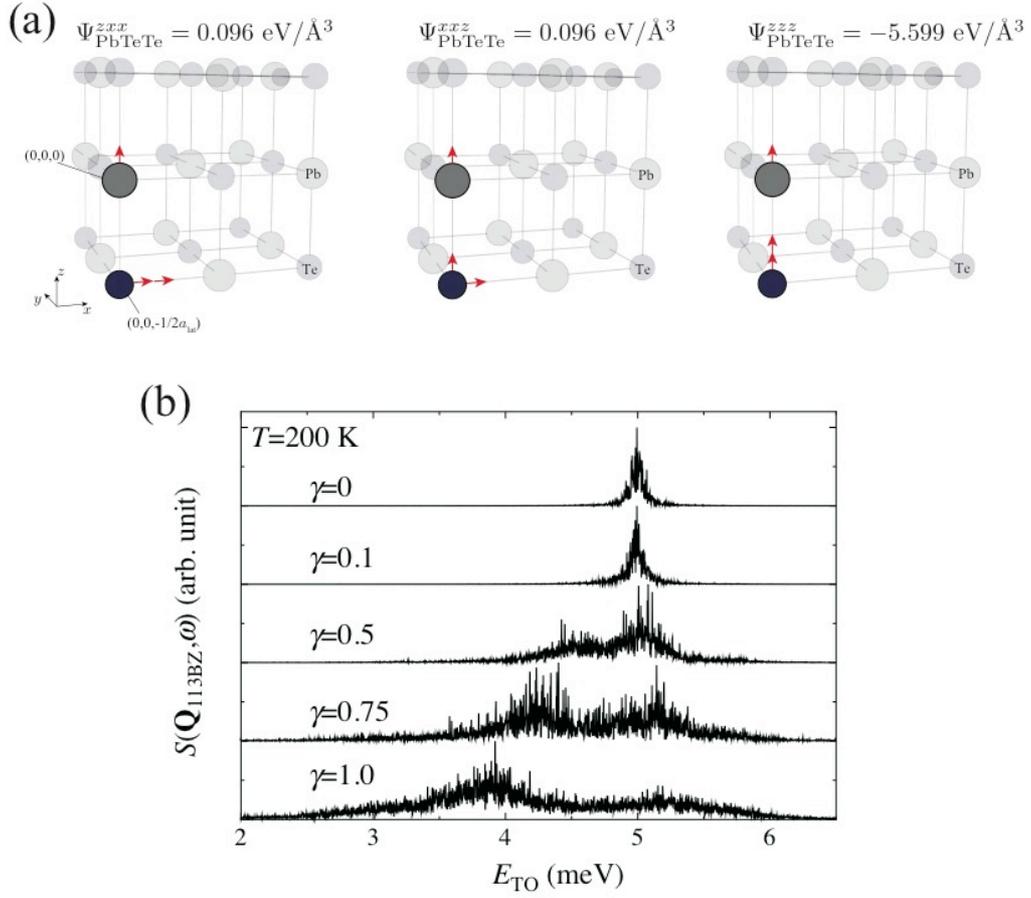

**Fig. 4.** (Color online) (a) Atomic displacements corresponding to three cubic anharmonic IFCs ($\Psi^{zxx}_{\text{PbTeTe}}$, $\Psi^{xxz}_{\text{PbTeTe}}$, and $\Psi^{zzz}_{\text{PbTeTe}}$) listed in Tab. 1. Atoms at (0,0,0) and (0, 0, -1/2)$a_{\text{lat}}$ are Pb and Te atoms, respectively. (b) $S(\mathbf{Q},\omega)$ spectra in (113) BZ at 200 K calculated with the cubic anharmonic IFCs along the [001] direction ($\Psi^{zzz}_{\text{PbTeTe}}$ and $\Psi^{zzz}_{\text{PbPbTe}}$) scaled by a factor $\gamma$.

**Tab. 1.** The irreducible nearest-neighbor cubic anharmonic IFCs of PbTe crystal ($\Psi_{ijk}^{\alpha\beta\gamma}$). Pb and Te atoms in this table are located at (0, 0, 0) and (0, 0, -1/2)$a_{\text{lat}}$, where $a_{\text{lat}}$ denotes the lattice constants ($a_{\text{lat}}$=6.548 Å). Integers *i*, *j*, and *k* are atomic indices, and *α*, *β*, and *γ* are Cartesian coordinates.

| # | *i* | *α* | *j* | *β* | *k* | *γ* | $\Psi_{ijk}^{\alpha\beta\gamma}$ (eV/ Å³) |
|---|---|---|---|---|---|---|---|
| 1 | Pb | *x* | Pb | *z* | Te | *x* | -0.096 |
| 2 | Pb | *x* | Pb | *x* | Te | *z* | -0.096 |
| 3 | Pb | *z* | Pb | *z* | Te | *z* | 5.599 |
| 4 | Pb | *z* | Te | *x* | Te | *x* | 0.096 |
| 5 | Pb | *x* | Te | *x* | Te | *z* | 0.096 |
| 6 | Pb | *z* | Te | *z* | Te | *z* | -5.599 |